\documentclass{elsart1p}
\usepackage{graphicx}
\usepackage{amssymb}

\begin{document}
\begin{frontmatter}
%
%
%
\title{Dynamic versus Static Hadronic Structure Functions\thanksref{label1}}
\thanks[label1]{This research was supported by the Department
of Energy  contract DE--AC02--76SF00515. SLAC-PUB-13507.}
%
%
\author{Stanley J. Brodsky}
\address{SLAC National Accelerator Laboratory, Stanford University, Stanford, CA 94309, USA}
\ead{sjbth@slac.stanford.edu}

\begin{abstract}
``Static" structure functions are the probabilistic distributions computed from the square of the light-front wavefunctions of the target hadron. In contrast, the ``dynamic"  structure functions measured in deep inelastic lepton-hadron scattering include the effects  of rescattering associated with the Wilson line.
Initial- and
final-state rescattering, neglected in the parton model, can have a profound effect in QCD hard-scattering reactions,
producing single-spin asymmetries, diffractive deep inelastic scattering, diffractive hard hadronic reactions, the breakdown of
the Lam-Tung relation in Drell-Yan reactions,  nuclear shadowing, and non-universal nuclear antishadowing---novel leading-twist physics not incorporated in
the light-front wavefunctions of the target computed in isolation.  I also review how ``direct" higher-twist processes -- where a proton is produced in the hard subprocess itself -- can explain the anomalous proton-to-pion ratio seen in
high centrality heavy ion collisions.  

\end{abstract}
\end{frontmatter}
%
\section{Introduction}
\begin{figure}[!]
\includegraphics[width=13cm]{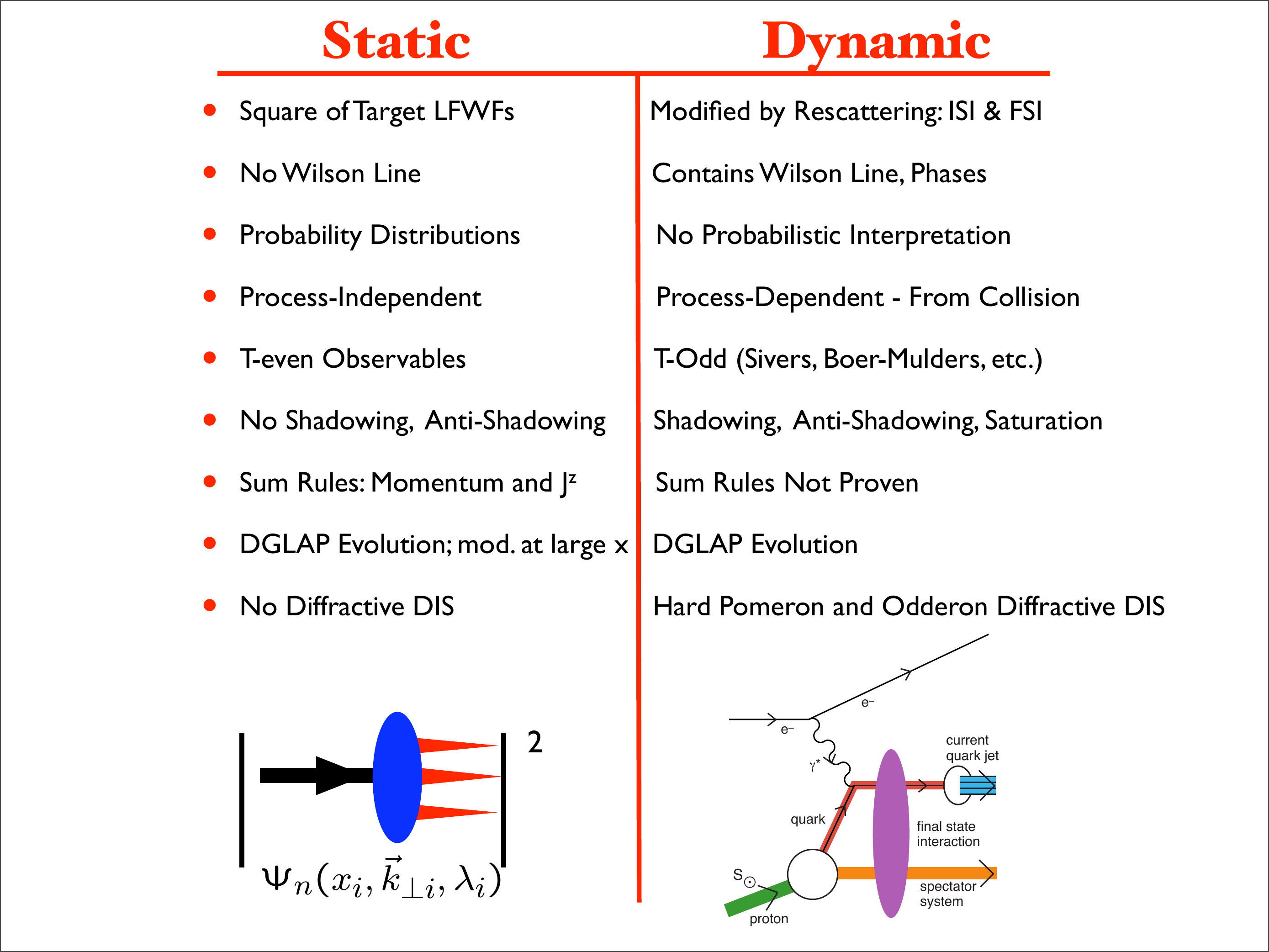}
\caption{Static versus dynamic structure functions}
\label{figstatdyn}  
\end{figure}

It is important to distinguish ``static" structure functions which are computed directly from the light-front wavefunctions of  a target hadron from the nonuniversal ``dynamic" empirical structure functions which take into account rescattering of the struck quark in deep inelastic lepton scattering.  [See  fig. \ref{figstatdyn}. ]
The real wavefunctions underlying static structure functions cannot describe diffractive deep inelastic scattering nor  single-spin asymmetries, since such phenomena involve the complex phase structure of the $\gamma^* p $ amplitude.   
One can augment the light-front wavefunctions with a gauge link corresponding to an external field
created by the virtual photon $q \bar q$ pair
current~\cite{Belitsky:2002sm,Collins:2004nx}, but such a gauge link is
process dependent~\cite{Collins:2002kn}, so the resulting augmented
wavefunctions are not universal.
\cite{Brodsky:2002ue,Belitsky:2002sm,Collins:2003fm}. 

A remarkable feature of deep inelastic lepton-proton scattering at
HERA is that approximately 10\% events are
diffractive~\cite{Adloff:1997sc,Breitweg:1998gc}: the target proton
remains intact, and there is a large rapidity gap between the proton
and the other hadrons in the final state.  
The presence of a rapidity gap
between the target and diffractive system requires that the target
remnant emerges in a color-singlet state; this is made possible in
any gauge by soft rescattering. The multiple scattering of the struck
parton via instantaneous interactions in the target generates
dominantly imaginary diffractive amplitudes, giving rise to an
effective ``hard pomeron'' exchange.   The resulting diffractive
contributions leave the target intact  and do not resolve its quark
structure; thus there are contributions to the DIS structure
functions which cannot be interpreted as parton
probabilities~\cite{Brodsky:2002ue}; the leading-twist contribution
to DIS  from rescattering of a quark in the target is thus a coherent
effect which is not included in the light-front wavefunctions
computed in isolation.

The shadowing of nuclear structure functions arises from 
destructive interference between multi-nucleon amplitudes involving
diffractive DIS and on-shell intermediate states with a complex
phase.  The physics of rescattering and nuclear shadowing is not
included in the nuclear light-front wavefunctions, and a
probabilistic interpretation of the nuclear DIS cross section is
precluded. 

Antishadowing of nuclear structure functions is also observed in deep
inelastic lepton-nucleus  scattering. Empirically, one finds $R_A(x,Q^2) \equiv  \left(F_{2A}(x,Q^2)/ (A/2) F_{d}(x,Q^2)\right)
> 1 $ in the domain $0.1 < x < 0.2$; {\em i.e.}, the measured nuclear structure function (referenced to the deuteron) is larger than than the
scattering on a set of $A$ independent nucleons.
Ivan Schmidt, Jian-Jun Yang, and I~\cite{Brodsky:2004qa} have extended the analysis of nuclear shadowing  to the shadowing and antishadowing of the
electroweak structure functions.  We note that there are leading-twist diffractive contributions $\gamma^* N_1 \to (q \bar q) N_1$  arising from Reggeon exchanges in the
$t$-channel~\cite{Brodsky:1989qz}.  For example, isospin--non-singlet $C=+$ Reggeons contribute to the difference of proton and neutron
structure functions, giving the characteristic Kuti-Weisskopf $F_{2p} - F_{2n} \sim x^{1-\alpha_R(0)} \sim x^{0.5}$ behavior at small $x$. The
$x$ dependence of the structure functions reflects the Regge behavior $\nu^{\alpha_R(0)} $ of the virtual Compton amplitude at fixed $Q^2$ and
$t=0.$ The phase of the diffractive amplitude is determined by analyticity and crossing to be proportional to $-1+ i$ for $\alpha_R=0.5,$ which
together with the phase from the Glauber cut, leads to {\it constructive} interference of the diffractive and nondiffractive multi-step nuclear
amplitudes.  The nuclear structure function is predicted to be enhanced precisely in the domain $0.1 < x <0.2$ where
antishadowing is empirically observed.  The strength of the Reggeon amplitudes is fixed by the fits to the nucleon structure functions, so there
is little model dependence.
Quarks of different flavors  will couple to different Reggeons; this leads to the remarkable prediction that
nuclear antishadowing is not universal; it depends on the quantum numbers of the struck quark. This picture implies substantially different
antishadowing for charged and neutral current reactions, thus affecting the extraction of the weak-mixing angle $\theta_W$.  We find that part
of the anomalous NuTeV result~\cite{Zeller:2001hh} for $\theta_W$ could be due to the non-universality of nuclear antishadowing for charged and
neutral currents.  In fact,  Schienbein et al.~\cite{Schienbein:2008ay} have recently given a comprehensive analysis of charged current deep inelastic neutrino-iron scattering, finding significant differences with the nuclear corrections for electron-iron scattering.

Diffractive multi-jet production in heavy nuclei provides a novel way to resolve the shape of light-front Fock state wavefunctions and test
color transparency~\cite{Brodsky:1988xz}.  For example, consider the reaction~\cite{Bertsch:1981py,Frankfurt:1999tq}.   $\pi A \rightarrow {\rm
Jet}_1 + {\rm Jet}_2 + A^\prime$ at high energy where the nucleus $A^\prime$ is left intact in its ground state. The transverse momenta of the
jets balance so that $ \vec k_{\perp i} + \vec k_{\perp 2} = \vec q_\perp < {R^{-1}}_A  .$  Because of color transparency, the valence wavefunction of the pion with small impact separation will penetrate the nucleus with minimal interactions, diffracting into jet
pairs~\cite{Bertsch:1981py}.  The $x_1=x$ and $x_2=1-x$ dependence of the dijet distributions thus reflects the shape of the pion valence
light-cone wavefunction in $x$; similarly, the $\vec k_{\perp 1}- \vec k_{\perp 2}$ relative transverse momenta of the jets gives key
information on the second transverse momentum derivative of the underlying shape of the valence pion
wavefunction~\cite{Frankfurt:1999tq}. The diffractive nuclear amplitude extrapolated to $t = 0$ will be linear in nuclear
number $A$ if color transparency is correct.  The integrated diffractive rate will then scale as $A^2/R^2_A \sim A^{4/3}$. This is in fact what
has been observed by the E791 collaboration at FermiLab for 500 GeV incident pions on nuclear targets~\cite{Aitala:2000hc}. 

\section{ Single-Spin Asymmetries  and Other Leading-Twist  Rescattering Effects}
Among the most interesting polarization effects are single-spin
azimuthal asymmetries  in semi-inclusive deep inelastic scattering,
representing the correlation of the spin of the proton target and
the virtual photon to hadron production plane: $\vec S_p \cdot \vec
q \times \vec p_H$. Such asymmetries are time-reversal odd, but they
can arise in QCD through phase differences in different spin
amplitudes. In fact, final-state interactions from gluon exchange
between the outgoing quarks and the target spectator system lead to
single-spin asymmetries (SSAs) in semi-inclusive deep inelastic
lepton-proton scattering  which  are not power-law suppressed at
large photon virtuality $Q^2$ at fixed
$x_{bj}$~\cite{Brodsky:2002cx}.  In contrast to the SSAs arising
from transversity and the Collins fragmentation function, the
fragmentation of the quark into hadrons is not necessary; one
predicts a correlation with the production plane of the quark jet
itself. Physically, the final-state interaction phase arises as the
infrared-finite difference of QCD Coulomb phases for hadron 
wavefunctions with differing orbital angular momentum.  The same proton
matrix element which determines the spin-orbit correlation $\vec S
\cdot \vec L$ also produces the anomalous magnetic moment of the
proton, the Pauli form factor, and the generalized parton
distribution $E$ which is measured in deeply virtual Compton
scattering. Thus the contribution of each quark current to the SSA
is proportional to the contribution $\kappa_{q/p}$ of that quark to
the proton target's anomalous magnetic moment $\kappa_p = \sum_q e_q
\kappa_{q/p}$ ~\cite{Brodsky:2002cx,Burkardt:2004vm}.  The SSA in the Drell-Yan
process is the same as that obtained in SIDIS, with the appropriate
identification of variables, but with the opposite sign.  If both the
quark and antiquark in the initial state of the Drell-Yan subprocess
$q \bar q \to  \mu^+ \mu^-$ interact with the spectators of the
other incident hadron, one finds a breakdown of the Lam-Tung
relation, which was formerly believed to be a general prediction of
leading-twist QCD. These double initial-state interactions also lead
to a $\cos 2 \phi$ planar correlation in unpolarized Drell-Yan
reactions~\cite{Boer:2002ju}.
As noted by Collins and Qiu~\cite{Collins:2007nk}, the traditional factorization formalism of perturbative QCD for high transverse
momentum hadron production  fails in detail even at the LHC because of initial- and final-state rescattering.  An important signal for factorization breakdown is a $\cos 2 \phi$ planar correlation in dijet production.

\section{Novel Intrinsic Heavy Quark Phenomena}
The probability for Fock states of a light hadron such as the proton to have an extra heavy quark pair decreases as $1/m^2_Q$ in non-Abelian
gauge theory~\cite{Franz:2000ee,Brodsky:1984nx}.  The relevant matrix element is the cube of the QCD field strength $G^3_{\mu \nu},$  in
contrast to QED where the relevant operator is $F^4_{\mu \nu}$ and the probability of intrinsic heavy leptons in an atomic
state is suppressed as $1/m^4_\ell.$  The maximum probability occurs at $x_i = { m^i_\perp /\sum^n_{j = 1}
m^j_\perp}$ where $m_{\perp i}= \sqrt{k^2_{\perp i} + m^2_i}.$; {\em i.e.}, when the constituents have minimal invariant mass and equal rapidity. Thus the heaviest constituents have the highest
momentum fractions and the highest $x_i$.  Intrinsic charm thus predicts that the charm structure function has support at large $x_{bj}$ in
excess of DGLAP extrapolations~\cite{Brodsky:1980pb}; this is in agreement with the EMC measurements~\cite{Harris:1995jx}. Intrinsic charm can
also explain the $J/\psi \to \rho \pi$ puzzle~\cite{Brodsky:1997fj}. It also affects the extraction of suppressed CKM matrix elements in $B$
decays~\cite{Brodsky:2001yt}.
The dissociation of the intrinsic charm $|uud c \bar c>$ Fock state of the proton can produce a leading heavy quarkonium state at
high $x_F = x_c + x_{\bar c}~$ in $p N \to J/\psi X A^\prime$ since the $c$ and $\bar c$ can readily coalesce into the charmonium state.  Since
the constituents of a given intrinsic heavy-quark Fock state tend to have the same rapidity, coalescence of multiple partons from the projectile
Fock state into charmed hadrons and mesons is also favored.  For example, one can produce a leading $\Lambda_c$ at high $x_F$ and low $p_T$ from
the coalescence of the $u d c$ constituents of the projectile $|uud c \bar c>$  Fock state. 
In the case of a nuclear target, the charmonium state will be produced at small transverse momentum and high $x_F$  with a characteristic
$A^{2/3}$ nuclear dependence since the color-octet color-octet $|(uud)_{8C} (c \bar c)_{8C} >$ Fock state interacts on the front surface of the
nuclear target~\cite{Brodsky:2006wb}. This forward contribution is in addition to the $A^1$ contribution derived from the usual perturbative QCD
fusion contribution at small $x_F.$   Because of these two components, the cross section violates perturbative QCD factorization for hard
inclusive reactions~\cite{Hoyer:1990us}.  This is consistent with the observed two-component cross section for charmonium production observed by
the NA3 collaboration at CERN~\cite{Badier:1981ci} and more recent experiments~\cite{Leitch:1999ea}. The diffractive dissociation of the
intrinsic charm Fock state leads to leading charm hadron production and fast charmonium production in agreement with
measurements~\cite{Anjos:2001jr}.  The hadroproduction cross sections for  double-charm $\Xi_{cc}^+$ baryons at SELEX~\cite{Ocherashvili:2004hi} and the production of $J/\psi$ pairs at NA3 are
be consistent with the diffractive dissociation and coalescence of double IC Fock states~\cite{Vogt:1995tf}. These observations provide
compelling evidence for the diffractive dissociation of complex off-shell Fock states of the projectile and contradict the traditional view that
sea quarks and gluons are always produced perturbatively via DGLAP evolution. It is also conceivable that the observations~\cite{Bari:1991ty} of
$\Lambda_b$ at high $x_F$ at the ISR in high energy $p p$  collisions could be due to the dissociation and coalescence of the
``intrinsic bottom" $|uud b \bar b>$ Fock states of the proton.
As emphasized by Lai, Tung, and Pumplin~\cite{Pumplin:2007wg}, there are indications that the structure functions used to model charm
and bottom quarks in the proton at large $x_{bj}$ have been  underestimated, since they ignore intrinsic heavy quark fluctuations of
hadron wavefunctions.   

Goldhaber, Kopeliovich, Schmidt, Soffer, and I ~\cite{Brodsky:2006wb,Brodsky:2007yz} have  proposed a novel  mechanism for exclusive diffractive
Higgs production $pp \to p H p $  and nondiffractive Higgs production  in which the Higgs boson carries a significant fraction of the projectile proton momentum. The production
mechanism is based on the subprocess $(Q \bar Q) g \to H $ where the $Q \bar Q$ in the $|uud Q \bar Q>$ intrinsic heavy quark Fock state has up
to $80\%$ of the projectile protons momentum. This mechanism provides a clear experimental signal 
for Higgs production at the LHC due to the small background in this kinematic region.

\section{Color Transparency and the RHIC Baryon Anomaly}
It is conventional to assume that leading-twist subprocesses dominate measurements of high $p_T$ hadron production at RHIC energies.  Indeed the
measured cross section for direct photon fragmentation 
${ E d\sigma/ d^3p}(p p \to \gamma X) ={F(x_T,\theta_{cm})/ p_T^{n_eff} } $ is consistent with $n_{eff}(p p \to \gamma X) = 5,$  as expected for the fixed-$x_T$ scaling of the $g q
\to \gamma q$ leading-twist subprocess~\cite{Brodsky:2005fza}.  However, the measured fixed-$x_T$ scaling for proton production at RHIC is anomalous:  PHENIX reports $n_{eff} (p p \to p X)\simeq 8$.  A review of this data is given by
Tannenbaum~\cite{Tannenbaum:2006ku}. One can understand the anomalous scaling if a higher-twist subprocess~\cite{Brodsky:2008qp} where the
proton is made {\it directly} within the hard reaction, such as $ u u \to p \bar d$ and $(uud) u \to p u$, dominates the reaction $ p p \to p X$
at RHIC energies.  
The dominance of direct subprocesses is possible since the fragmentation of gluon or quark jets
to baryons requires that the  2 to 2 subprocess occurs at much higher transverse momentum than the $p_T$ of observed proton because of the fast-falling $(1-z)^3 $ quark-to-proton fragmentation function.   Thus the initial quark and gluon distributions have to be evaluated at higher $x$ in leading twist fragmentation reactions compared to direct processes.
Such ``direct" reactions can readily explain the fast-falling power-law  falloff
observed at fixed $x_T$ and fixed-$\theta_{cm}$ observed at the ISR, FermiLab and RHIC.
Furthermore, the protons produced
directly within the hard subprocess emerge as small-size color-transparent colored states which are not absorbed in the nuclear target. In
contrast, pions produced from jet fragmentation have the normal cross section. This provides a plausible explanation of the RHIC
data,~\cite{Adler:2003kg} which shows a dramatic rise of the $p / \pi$ ratio at high $p_T$ and a higher value for $n_{eff}$ at fixed $x_T$ when one compares  peripheral  with  central (full
overlap) heavy ion collisions.  The directly produced protons are not absorbed, but the pions are
diminished in the nuclear medium.


%
%
%

%
\end{document}